# The hyperfine Paschen–Back Faraday effect

Mark A Zentile[1], Rebecca Andrews[1], Lee Weller[1], Svenja Knappe[2], Charles S Adams[1] and Ifan G Hughes[1]

[1] Joint Quantum Center (JQC) Durham-Newcastle, Department of Physics, Durham University, South Road, Durham, DH1 3LE, UK
[2] Time and Frequency Division, The National Institute of Standards and Technology, Boulder, CO 80305, USA

E-mail: m.a.zentile@durham.ac.uk

## Abstract
We investigate experimentally and theoretically the Faraday effect in an atomic medium in the hyperfine Paschen–Back regime, where the Zeeman interaction is larger than the hyperfine splitting. We use a small permanent magnet and a micro-fabricated vapour cell, giving magnetic fields of the order of a tesla. We show that for low absorption and small rotation angles, the refractive index is well approximated by the Faraday rotation signal, giving a simple way to measure the atomic refractive index. Fitting to the atomic spectra, we achieve magnetic field sensitivity at the $10^{-4}$ level. Finally we note that the Faraday signal shows zero crossings which can be used as temperature insensitive error signals for laser frequency stabilization at large detuning. The theoretical sensitivity for $^{87}$Rb is found to be $\sim 40$ kHz $°C^{-1}$.

Keywords: Faraday effect, Zeeman shift, hyperfine splitting, magneto-optical effects, dispersion

(Some figures may appear in colour only in the online journal)

## 1. Introduction

Thermal atomic vapours are finding an ever-increasing array of applications, providing high sensitivity in relatively simple experiments. Chip-scale atomic clocks [1] and magnetometers [2], quantum memories [3], microwave electric field detection [4], microwave magnetic field imaging [5, 6], frequency filtering [7–10], optical isolation [11], high-bandwidth measurement [12], laser frequency stabilising [13, 14], orbital angular momentum transfer [15], measuring number density in an optically thick medium [16] and creating a medium with a giant optical nonlinearity [17] have all been demonstrated.

Many of these applications rely on the Faraday effect which arises due to a magnetic field producing circular birefringence in the medium. Some applications require very large fields [10, 11]. In recent years these large magnetic fields of the order of one tesla have become more easily achievable owing to the availability of inexpensive neodymium-based permanent magnets. At these large fields alkali-metal atoms enter the hyperfine Paschen–Back (HPB) regime [18–23], where the nuclear, $I$, and the total electronic, $J$, angular momenta are decoupled. Exceptionally, for lithium a magnetic field of the order of a tesla is strong enough to also decouple the orbital, $L$ and spin, $S$, angular momentum of the electron [24]. The HPB regime is also of interest for coherent dynamics as individual transitions become separately addressable [25].

The Faraday effect in very strong magnetic fields has been studied before in the case of Rydberg transitions [26, 27]. However, to the best of our knowledge, it has not been studied for the main resonance lines (D lines).

In this paper we compare theory and experiment for the Faraday effect of an atomic vapour in the HPB regime, as defined in section 2. This extends previous work on absolute absorption [28] and dispersion [29] at high densities [30, 31] and high magnetic fields [32].

The structure of the paper is as follows: section 2 describes the theory used to model the experiment. Section 3 shows that under suitable conditions Faraday rotation can be used to measure the refractive index of the medium directly. In section 4 we describe the experimental apparatus and show the spectra obtained. We compare the spectra to theory and show that there is excellent agreement, also noting that the relative sensitivity to the average magnetic field is at the $10^{-4}$ level. In section 5 we show that the Faraday spectra obtained can be used as highly temperature insensitive laser frequency references, which are far-detuned from the zero-field resonances. Finally, we draw our conclusions in section 6.

## 2. Theory

The HPB regime is defined when the Zeeman interaction is larger than the hyperfine splitting. However, The HPB Faraday effect occurs when the Zeeman shift, $\Delta_Z$, is larger than the Doppler width, $\Delta_D$. It is distinct from the resonant Faraday effect [33] where the laser detuning, $\Delta$, is in the regime $|\Delta| < \Delta_D \approx |\Delta_Z|$, or the off-resonant Faraday effect [34] where $|\Delta| > \Delta_D \approx |\Delta_Z|$. In our experiment we use $^{87}$Rb on the D$_2$ line ($5^2S_{1/2} \rightarrow 5^2P_{3/2}$) to investigate the effect experimentally. To model the effect we calculate both the absorption and dispersion of light by the atomic vapour.

The (complex) index of refraction, $n_c = n + i\beta$, allows one to calculate the absorptive and dispersive properties of an optical medium [35]. The real part, $n$, is the ratio of the speed of light in vacuum, $c$, to the phase velocity in the medium. The imaginary part, $\beta$, gives the extinction coefficient, $\alpha$, using the equation $\alpha = 2k\beta$ [35], where $k$ is the magnitude of the wave vector. The theoretical model we have used to calculate the index of refraction as a function of frequency is described in detail in [28–30, 32]. Here we briefly summarise key points.

### 2.1. The electric susceptibility

The index of refraction is related to the electric susceptibility $\chi$ by $n_c = \sqrt{1+\chi}$ [35]. For an ensemble of two-level atoms, the electric susceptibility is calculated by multiplying the transition strength factor by the number density and a line-shape profile (a complex value) [28]. The real part of the susceptibility has a characteristic dispersion profile. The imaginary part has a Voigt profile arising from homogeneous broadening (Lorentzian) and Doppler broadening (Gaussian). The full width at half maximum of the Lorentzian broadening ($\Gamma$) has contributions from natural broadening ($\Gamma_0$), buffer gases ($\Gamma_{\text{buffer}}$) and dipole–dipole induced self-broadening ($\Gamma_{\text{self}}$) [30].

The full weak-probe susceptibility for an ensemble of real multi-level atoms can be found by summing the contribution for each transition, taking into account the different line-strengths and line-centre frequencies.

### 2.2. The atomic Hamiltonian

In order to find the transition frequencies and strengths, we construct an atomic Hamiltonian that includes both hyperfine and external magnetic-field [29] interactions in the uncoupled basis. In previous work [11, 29, 32] we neglected the contribution from the magnetic moment of the nucleus. However, for the required precision of this study, we include the nuclear magnetic moment. The coarse atomic energy and fine interaction are chosen such that zero linear detuning, $\Delta/2\pi$, occurs at the weighted line-centre of naturally abundant rubidium at zero magnetic field (384.230 4266 THz). Expressing the atomic Hamiltonian as a matrix allows the eigenstates, eigenvalues and transition strengths to be calculated easily by numerical methods.

**Table 1.** Good quantum numbers characterizing the $\sigma^-$ (negatively detuned) HPB transitions 1–8, as labelled in figures 3 and 4. $m_J$ and $m_J'$ refer to the projection on the magnetic field axis of the total electronic angular momentum $J$ in the ground and excited state respectively. Also given is the shift of the transition line-centre with a magnetic field around 0.55 T, as calculated with the Hamiltonian model described in section 2.2 and by the approximation of equation (1).

| | | $\partial \nu_0 / \partial B$ (MHz/gauss) | | | |
|---|---|---|---|---|---|
| Transition | $m_I$ | Hamiltonian | Approximation | $m_J$ | $m_J'$ |
| 1 | 3/2 | −1.46 | | | |
| 2 | 1/2 | −1.52 | −1.40 | $-\frac{1}{2}$ | $-\frac{3}{2}$ |
| 3 | −1/2 | −1.54 | | | |
| 4 | −3/2 | −1.40 | | | |
| 5 | −3/2 | −2.19 | | | |
| 6 | −1/2 | −2.21 | −2.33 | $+\frac{1}{2}$ | $-\frac{1}{2}$ |
| 7 | 1/2 | −2.27 | | | |
| 8 | 3/2 | −2.33 | | | |

It should be noted that we have only considered the case where the magnetic field direction is parallel to the propagation of the light. Thus we have imposed the selection rules [36] $\Delta m_I = 0$, $\Delta m_S = 0$ and $\Delta m_L = \pm 1$, where $m_I$, $m_S$ and $m_L$ are the projections of the nuclear spin, electron spin and electron orbital angular momentum on the magnetic field axis. These are known as $\sigma^\pm$ ($\Delta m_L = \pm 1$) transitions. The presence of the magnetic field causes circular birefringence and dichroism since there is a different index of refraction, $n_c^\pm$, for each hand of light.

### 2.3. Approximation for energy level shifts in the hyperfine Paschen–Back regime

When the hyperfine interaction is small compared to the magnetic field interaction we can approximate the shifts in the energy levels ($\Delta E$) using [37]

$$\Delta E \approx (g_J m_J \mu_B + g_I m_I \mu_N) B, \quad (1)$$

where $g_J$ and $g_I$ are the gyromagnetic ratios corresponding to $J$ and $I$, $\mu_B$ is the Bohr magneton, $\mu_N$ is the nuclear magneton and $B$ is the magnitude of the applied magnetic field. The values of $g_I$ for $^{87}$Rb and $^{85}$Rb were taken to be −1.827 2315(18) and −0.539 168(1) respectively [38]. This equation can be useful to quickly estimate line-shifts with changes in magnetic field, since the line shift is given by the difference of the excited-level shift to the ground-level shift. Table 1 shows a comparison of this approximation to the more accurate Hamiltonian method at the magnetic field of our experiment.

## 3. Faraday rotation as a direct measure of refractive index

The Faraday effect has already been shown to be a good direct measure of absolute differential dispersion ($n^+ - n^-$) [7, 34].

Measuring a spectrum (e.g. transmission) and fitting to the comprehensive model described in sections 2.1 and 2.2 allows the individual refractive indices, $n^+$ and $n^-$ to be extracted indirectly. In general Faraday rotation cannot be used as a direct measure since both $n^+$ and $n^-$ tend to change with detuning; other techniques need to be used [39, 40]. However, we show that measuring Faraday rotation in the HPB regime gives a good approximation to the individual refractive indices of the medium.

The output polarization is conveniently parametrized in terms of the Stokes parameter [41] $S_1$, defined as $S_1 \equiv (I_x - I_y)/I_0$, where $I_x$ and $I_y$ are the intensity of linear polarization components in the orthogonal $x$ and $y$ axes, and $I_0$ is the total intensity before the light traverses the medium. For linearly polarized light incident on the medium, $S_1$ is calculated using the equation [32]

$$S_1 = \cos(2\psi) \exp\left[-\tfrac{1}{2}(\alpha^- + \alpha^+)L\right], \quad (2)$$

where $\psi$ is the rotation angle with respect to the $x$-axis, $\alpha^\pm$ are the extinction coefficients of the hand of light that drives $\sigma^\pm$ transitions and $L$ is the length of the medium. The rotation angle is simply given by $\psi = \tfrac{1}{2}(\phi^+ - \phi^-) + \theta_0$, where $\theta_0$ is the initial angle, and $\phi^\pm = kn^\pm L$ are the phase shifts of the hand of light that drives $\sigma^\pm$ transitions. Performing a Taylor-series expansion of the cosine (around $2\theta_0$) and exponential parts of equation (2), and writing explicitly in terms of the real and imaginary parts of $n_c^+$ and $n_c^-$ gives

$$\cos(2\psi) = \cos(2\theta_0) - kL(n^+ - n^-)\sin(2\theta_0)$$
$$- \tfrac{1}{2}[kL(n^+ - n^-)]^2 \cos(2\theta_0) + \cdots, \quad (3)$$

$$\exp[-kL(\beta^+ - \beta^-)] = 1 + kL(\beta^+ - \beta^-)$$
$$+ \tfrac{1}{2}[kL(\beta^+ - \beta^-)]^2 + \cdots. \quad (4)$$

*Condition 1.* We assume that the $\sigma^+$ and $\sigma^-$ transitions are far from each other in frequency space, which is the case when in the HPB regime. If we consider the part of the spectrum far from $\sigma^+$ transitions then $n^+ \to 1$ and $\beta^+ \to 0$.

*Condition 2.* The optical depth of the medium is small implying that $kL|(n^- - 1)|$ and $kL\beta^-$ are small, such that we can neglect terms higher than first order in the Taylor-series expansions. Note that by inspection of equation (3) this also requires $\theta_0$ be set close to $45°$ (balanced polarimetry [33]), otherwise the spectrum will be insensitive to rotation. This leaves us with

$$S_1 \approx (n^- - 1)kL(1 + \beta^- kL). \quad (5)$$

*Condition 3.* If we further restrict $\beta^-$ such that $\beta^- kL \ll 1$ then we derive our final result,

$$S_1 \approx (n^- - 1)kL. \quad (6)$$

If we instead consider the spectrum far from $\sigma^-$ transitions, we change condition 2 such that $n^- \to 1$ and $\beta^- \to 0$. Repeating the derivation would give us the complimentary result,

$$S_1 \approx (1 - n^+)kL. \quad (7)$$

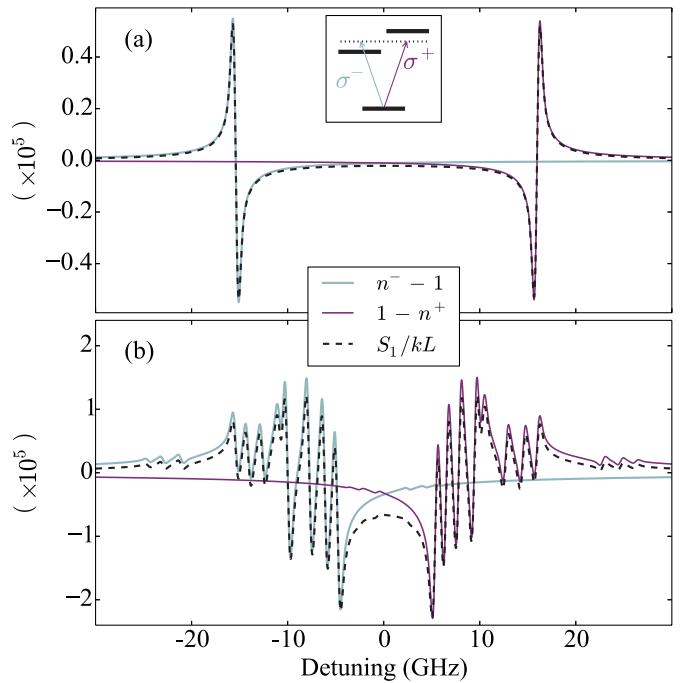

**Figure 1.** Faraday rotation signal $S_1/kL$ and circular refractive indices plotted against detuning. Panel (a) is modelled for a single $\sigma^+$ and $\sigma^-$ transition each. Panel (b) is modelled for $^{87}$Rb in a 1 mm long cell at a magnetic field of 0.55 T and a temperature of 70 °C (this fixes both the Doppler width and the number density [42]). The dashed black line corresponds to $S_1/kL$ while the light blue and purple lines correspond to $n^- - 1$ and $1 - n^+$ respectively.

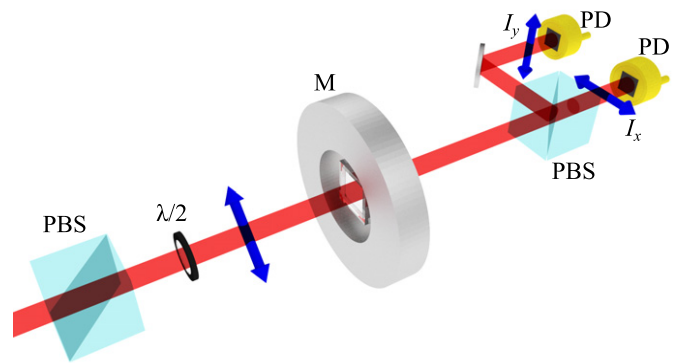

**Figure 2.** Schematic of the experimental set-up. PBS = polarizing beam splitter cube, $\lambda/2$ = half-wave plate, M = permanent magnet, PD = photodiode. Linearly polarized light set at an angle of $45°$ is incident on the thermal vapour cell. After the cell the horizontal and vertical polarizations of the beam are separated and measured on separate PD detectors.

Note that the validity of equations (6) and (7) improves for larger magnetic fields, since condition 1 will be better satisfied. Figure 1 plots $S_1/kL$ along with $n^- - 1$ and $1 - n^+$ against detuning. $S_1$ was calculated without approximation using equation (2). In panel (a) we have modelled a simple system of three-level atoms with a single $\sigma^+$ and $\sigma^-$ transition far detuned from each other. $S_1/kL$ gives an excellent approximation to the refractive indices when close to their respective transitions. Panel (b) shows the model of $^{87}$Rb vapour on the D$_2$ line at 0.55 T. We can see that $S_1/kL$ gives a reasonable approximation to $n^- - 1$ and $1 - n^+$, except in the region around zero global detuning, where the frequency of light is not sufficiently far from resonance for either refractive index to be close to 1. We now investigate such Faraday rotation spectra experimentally.

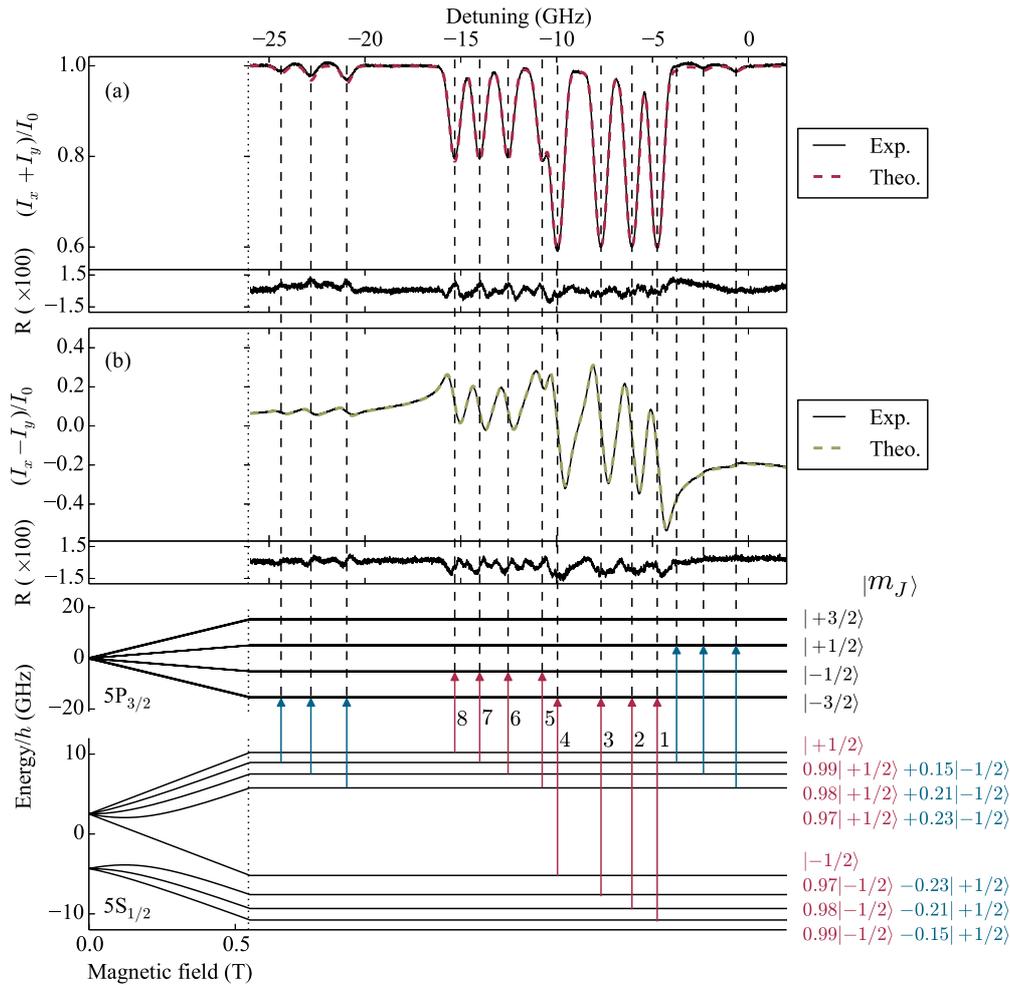

**Figure 3.** Experimental and theoretical transmission and Faraday rotation spectra for the 1 mm long Rb vapour cell (99% $^{87}$Rb, 1% $^{85}$Rb) with rubidium vapour number density of $\mathcal{N} = 3.10 \times 10^{12}$ cm$^{-3}$. Panel (a) shows the sum of the horizontal ($I_x$) and vertical ($I_y$) components of the probe beam divided by the initial intensity ($I_0$), equivalent to transmission. The solid black curve shows the experimental data, while the dashed red curve is the result of a fit to the model. Panel (b) shows $(I_x - I_y)/I_0$, the Stokes parameter $S_1$, using the same experimental data as that in panel (a). The dashed olive line is an independent fit to the model. The bottom sections of panels (a) and (b) show the residuals (R). We see excellent agreement between experiment and theory with rms deviations of 0.5% and 0.7% for the transmission and Faraday rotation spectra receptively. From fitting the transmission spectrum, the magnetic field was found to be $(0.5453 \pm 0.0002)$ T, the cell temperature was $(89.95 \pm 0.05)$ °C, the buffer gas broadening was $\Gamma_{\text{buffer}}/2\pi = (47 \pm 2)$ MHz. From fitting the $S_1$ spectrum, the magnetic field was $(0.5454 \pm 0.0002)$ T, the temperature was $(89.77 \pm 0.07)$ °C, the buffer gas broadening was $(60.6 \pm 0.3)$ MHz and the initial polarization angle was $(44.515 \pm 0.006)°$. Below panel (b) the energy levels are shown with their eigenstates (labelled by $m_J$) calculated at 0.55 T. The excited state manifold shows 16 levels in four groups of 4 but the spacings between levels in a group are too small to be resolved in the diagram. The transitions labelled 1–8 correspond to table 1. Note that the transitions are $\sigma^-$ except for the three weakly allowed transitions on the right, which are $\sigma^+$.

## 4. Experimental apparatus and results

The experimental apparatus used was similar to that used in [11, 32]. Figure 2 shows a schematic of the experimental set-up. An external cavity diode laser was scanned ∼30 GHz around the rubidium D$_2$ transition. Due to the limited mode-hop free tuning range of the laser, we restricted the scan to the negatively detuned part of the spectrum. A Fabry–Perot etalon and saturated absorption/hyperfine pumping spectroscopy [43, 44] (not shown in figure 2) was used to calibrate the laser scan [28]. The beam was then attenuated by a neutral-density filter ensuring weak-probe regime [45] was achieved. The beam then travelled through a lens of focal length 30 cm, a polarizing beam splitter (PBS) cube and a half-wave plate ($\lambda/2$) before being focussed onto the experimental cell. The $1 \times 1 \times 1$ mm$^3$ micro-fabricated vapour cell [46] was held within a permanent magnet. The cell was placed in a copper cradle, which was heated by passing a current through a ceramic resistor (see [47] for further details). After the cell the beam was collimated using another lens before being split into its horizontal ($I_x$) and vertical ($I_y$) polarizations with a PBS. These two beams were then measured on separate photodiodes (PD).

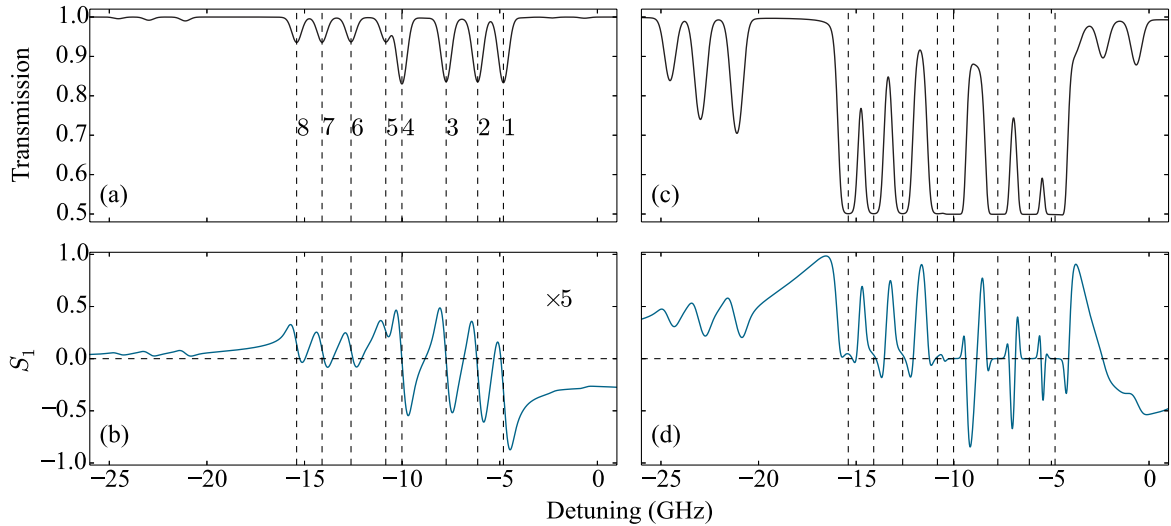

**Figure 4.** Theoretical transmission and $S_1$ spectra through a 1 mm long rubidium vapour (99% $^{87}$Rb, 1% $^{85}$Rb) with an applied magnetic field of 0.55 T. Panels (a) and (b) correspond to a vapour temperature of 70 °C, while panels (c) and (d) correspond to 130 °C. The data in panel (b) have been multiplied by a factor of 5 for clarity. Transitions labelled 1–8 correspond to table 1. These transitions are the $\sigma^-$ (negatively detuned) HPB transitions for which $m_J$ and $m_I$ are good quantum numbers. For the lower temperature case we can see that the transitions 1–4 are three times as strong as the transitions 5–8, as expected due to the $m_J = 1/2 \to m'_J = -1/2$ transitions sharing their strength with the D$_1$ line [47]. However, in the higher-temperature case, the increase in number density has caused the medium to be optically thick at some frequencies for one circular polarization. Also, at higher temperatures the $S_1$ spectrum (d) no longer approximates the refractive index since both absorption and rotation are large.

Figure 3 shows an example transmission and $S_1$ signal measured for a single laser scan. Excellent agreement between experiment and theory can be seen. Also shown in figure 3 are the atomic energy levels labelled by their eigenstates in the $m_J$ basis ($m_I$ labels are not included). The diagram of the energy levels shows that the weak transitions arise from small components of the eigenstates giving allowed $\Delta m_J = \pm 1$ transitions. It should be noted that the eight strong HPB transitions are insensitive to the nuclear magnetic moment while the weak transitions are sensitive. This can be seen from equation (1) and recalling the $\Delta m_I = 0$ selection rule. The fact that the energy of the eigenstates are mostly governed by the largest contribution to the state, necessarily implies this difference in sensitivity.

Five data sets were taken in quick succession and each was fitted to the model. We used a method similar to random-restart hill climbing [48] to avoid fitting to a local minimum, with the Marquardt–Levenberg method used to perform $\chi^2$ minimization and find the optimum parameters [49]. See the caption in figure 3 for the values of the fit parameters with their standard errors. These statistical uncertainties were found to be very small; systematic uncertainties are likely to dominate. See the appendix for further details on the systematic uncertainties.

The value of the magnetic field we extract corresponds to the average magnetic field across the beam path. By measuring the field profile produced by our magnet [32], we estimated a ~2% magnetic field variation across the beam path. We achieve a small fractional statistical uncertainty in the average magnetic field of $4 \times 10^{-4}$.

## 5. Laser frequency stabilization at large detuning

It has already been shown by Marchant *et al* [50] that a Faraday rotation signal can be used as a laser frequency reference at large values of detuning. In that system, with relatively low magnetic fields, zero crossings are achieved when the Faraday rotation is a multiple of 90°, but not at zero rotation. Since all the atoms in the path of the laser beam take part in rotating the plane of polarization, these zero crossings are sensitive to number density. Since number density is a near exponential function with temperature [42], the frequency where the zero crossings occur change rapidly with temperature (~0.2 GHz °C$^{-1}$ [50]). Therefore, the cell temperature must be controlled carefully and may require active stabilization.

In contrast, figure 3 shows an $S_1$ spectrum with zero crossings that occur for zero rotation angle. They occur due to the fact that $n^- - 1$ changes sign when the detuning is scanned over a $\sigma^-$ resonance, while the refractive index for the light driving $\sigma^+$ remains constant at approximately 1. This also happens between resonances. These changes in speed across $c$ cause a change in rotation direction and hence the zero crossings occur when the rotation is zero (when both circular polarizations of the light travel at $c$). This indicates that the position of these zero crossings should be temperature insensitive. Theoretically, the limiting factors to the temperature stability are the rotation caused by nearby resonances and the small deviation of $n^+$ from 1.

The effect was characterized theoretically by changing the cell temperature from 65 to 110 °C with all other parameters fixed ($B = 0.55$ T, $\theta_0 = 45°$, $\Gamma_{\text{buffer}} = 23.7$ MHz). This showed that the zero crossings typically move by less than 40 kHz °C$^{-1}$.

However, in real experimental conditions there will be more limitations. For example any offset in the signal, which for example could be caused by deviations of the input polarization from 45°, means that the zero crossing will no longer occur exactly at zero rotation. Other limitations arise from correlations between the other parameters and cell temperature. For example an increase in cell temperature is likely to cause heating of the surrounding neodymium magnet, which in turn will cause reversible demagnetization [51]. To investigate these effects experimentally, transmission and $S_1$ spectra were taken at cell temperatures ranging from 60 to 125 °C. By fitting the spectra, a linear correlation between the cell temperature and the magnetic field was found with a gradient of $(-2.21 \pm 0.03)$ gauss °C$^{-1}$. From the manufacturer's specifications we expect ∼7 gauss change in magnetic field per degree change in magnet temperature. This shows that the magnet temperature increased by about 0.3 °C per one degree increase in cell temperature. It should be noted that there was no attempt to insulate the permanent magnet from the cell heater [52]. Despite this the zero crossings were found to move by no more than ∼ 5 MHz per one degree change in cell temperature.

At high cell temperatures the medium starts to become optically thick for one circular polarization. When this happens the sharp zero crossings seen directly on resonance in the $S_1$ spectrum disappear. Figure 4 shows the effect on the atomic spectra when moving from the low-density regime to the high-density/optically-thick regime. However, the zero crossings between resonance can still be seen, with additional ones forming as a result of Faraday rotation of 90°. It is worth noting that the zero crossing between transitions 3 and 4 becomes a good reference at this high optical depth. Figure 4 clearly shows that when increasing the temperature of the vapour cell the amplitude of the signals arising from the three weak far detuned transitions (at −25 to −20 GHz) become significant. However, the Faraday rotation induced from these resonances is always weaker than the off-resonant rotation of the stronger HPB transitions. This means that one cannot use the weak resonances to make zero crossings that occur at zero rotation. Therefore, these weak transitions cannot be used as highly temperature stable frequency references, in contrast to the HPB transitions.

## 6. Conclusions

We have investigated the Faraday effect in an atomic medium in the hyperfine Paschen–Back regime. We have shown that in this regime a Faraday spectrum can be used measure directly the refractive index of the vapour. We have chosen $^{87}$Rb to model the effect both experimentally and theoretically, and have found excellent agreement between theory and experiment. We have demonstrated sensitivity to fields of order one tesla at the $10^{-4}$ level. We have also shown that the Faraday spectra have zero crossings that can be used to make laser-frequency error signals which are temperature insensitive, with a theoretical stability on the order of tens of kHz °C$^{-1}$.

## Acknowledgments

We acknowledge financial support from EPSRC (grants EP/H002839/1 and EP/F025459/1) and Durham University. The data presented in this paper are available upon request. RA was funded by the Ogden trust.

## Appendix. Experimental systematic and statistical uncertainties

The systematic uncertainty in magnetic field was found by estimating shifts due to error in laser scan calibration (∼10 MHz) and buffer gasses (∼8 MHz [32]), giving a value of ∼10$^{-3}$ T. The systematic uncertainty in the temperature was found to be 0.7 °C, given by the uncertainty in the vapour pressure formula used [42]. For the initial polarization angle, the uncertainty in the difference between the collection and detection on the two PD was estimated and equation (3) was used to turn this into an uncertainty in angle. This was found to be approximately 0.3 °. The transmission spectrum was found to give a value of $\Gamma_{\text{buffer}}$ that was ∼23 MHz larger than the $(23.7 \pm 1.2)$ MHz previously found from fitting spectra with zero applied magnetic field [32]. We attribute this to non-uniformity of the magnetic field across the path length of the beam. Simulating the non-uniform magnetic field was done by averaging many theoretical transmission spectra over a range of field values. Fitting to this average spectrum gave a value of $\Gamma_{\text{buffer}}$ that was ∼20 MHz larger for a 2.5% field variation. Also, the value of $\Gamma_{\text{buffer}}$ found from fitting to the $S_1$ spectrum was found to give a further ∼15 MHz larger value which, from theoretical simulations, were likely to be caused by imperfections in the polarising optics after the cell. These imperfections will affect the $S_1$ signal but not the transmission spectrum.

The statistical uncertainty in the experiment is largely attributed to detector noise. If desired, further improvements could be made by modulating the input probe beam and then using a lock-in amplifier to recover the signal [53].